\documentclass[conference]{IEEEtran}
\usepackage{cite}
\usepackage{amsmath,amssymb,amsfonts}
\usepackage{algorithmic}
\usepackage{graphicx}
\usepackage{textcomp}
\usepackage{xcolor}
\def\BibTeX{{\rm B\kern-.05em{\sc i\kern-.025em b}\kern-.08em
    T\kern-.1667em\lower.7ex\hbox{E}\kern-.125emX}}
\begin{document}

\title{A Distributed Computing Framework for Satellite Swarms}

\author{\IEEEauthorblockN{Ezra Fielding\IEEEauthorrefmark{1}, Clement Demazure\IEEEauthorrefmark{2}, Guthemberg Silvestre\IEEEauthorrefmark{3}, Felipe Alves Suana\IEEEauthorrefmark{1}, Philippe Queinnec\IEEEauthorrefmark{2}}
\IEEEauthorblockA{\IEEEauthorrefmark{1}CNES, Toulouse, France\\
ezra.fielding@cnes.fr, felipe.alvessuana@cnes.fr}
\IEEEauthorblockA{\IEEEauthorrefmark{2}Toulouse INP -- IRIT, Toulouse, France \\
clement.demazure@etu.toulouse-inp.fr, philippe.queinnec@irit.fr}
\IEEEauthorblockA{\IEEEauthorrefmark{3}Fédération ENAC ISAE-SUPAERO ONERA, Université de Toulouse, France \\
silvestre@enac.fr}
}

\maketitle

\begin{abstract}
The rise of large satellite constellations and Distributed Space Systems (DSS) demands generalized frameworks that enable fault-tolerant, autonomous distributed space applications. Conventional ground-centric command and control does not scale to systems of tens or hundreds of satellites, motivating the adoption of distributed computing. This paper introduces a conceptual distributed computing framework for satellite swarms, covering distributed state, command and control, and scientific mission. As a first validation step, a strongly eventually consistent distributed state service is designed and implemented using Conflict-free Replicated Data Types (CRDT), specifically a Last-Write-Wins Register-based key-value store. The service is evaluated in the context of Space Situational Awareness catalog dissemination across a simulated 66-satellite constellation emulated with GoNetEm. Results show that the CRDT-based approach reduces ground-to-satellite communication from 66 messages to a single uplink per update, with tree traversal matching the total message count of direct uplink. Under concurrent updates, sub-linear inter-satellite message growth is observed as nodes discard outdated versions, further reducing network overhead. These results demonstrate the viability of a CRDT-based distributed state as a scalable and fault-tolerant foundation for satellite swarm applications.
\end{abstract}


\section{Introduction}
\label{sec:introduction}
Improved access to space and advances in technology have sparked a shift in how space missions are executed. Large numbers of satellites can now be cheaply built and sent to space. At the end of 2024, 13672 payload objects were orbiting the Earth~\cite{esa_space_environment_2026}. This number will only go up as launch cadences increase to facilitate the emergence of large satellite constellations composed of tens, hundreds or even thousands of satellites. For the benefits of this increase to be felt, an efficient and scalable means is required to command and control these large, distributed space platforms. The large number of entities in these platforms coupled with the constraints imposed by the harsh space environment forces one to consider novel approaches to what has historically been implemented in space.

These developments have accelerated the shift towards the implementation of Multi-Spacecraft Systems (MSS), also known as Distributed Space Missions (DSM), where many satellites collaborate to complete the same mission~\cite{adamsOverviewDistributedSpacecraft2023, moigneNewObservingStrategy2019}. MSS sits in contrast to conventional monolithic space missions, where a single spacecraft incrementally completes mission objectives~\cite{adamsOverviewDistributedSpacecraft2023}. Conventional mission planning, operations, and data handling will not scale in large multi-spacecraft environments, resulting in an increased burden on human operators and ground infrastructure \cite{cellucciDistributedSpacecraftAutonomy2020}. The increased burden prevents the benefits of MSS from being fully realized. To overcome this burden, control and decision authority can be moved from the ground to the satellites themselves, shifting satellite constellations to the concept of satellite swarms. Satellite swarms fall under the umbrella of Automated Multi-Spacecraft Systems (A-MSS) and Distributed Space Systems (DSS)~\cite{adamsOverviewDistributedSpacecraft2023}. DSS can be considered a subset of A-MSS, with A-MSS itself a subset of MSS. The main distinction between A-MSS and DSS lies in how control authority is distributed. In DSS, control authority is distributed across all spacecraft in the system, resulting in a number of advantages, such as improved scalability, fault-tolerance and redundancy~\cite{adamsOverviewDistributedSpacecraft2023}. This makes DSS the more challenging problem to address, while, at the same time, making it most ideal for enabling fault-tolerant and autonomous satellite swarms which can be operated and interacted with as a single entity.

A number of works have begun the investigation into making DSS a reality. The findings of these studies will inform the approaches taken for future DSS missions, however, the research largely focuses on mission specific implementations and fails to address the need for a generalized framework to enable DSS and satellite swarms more broadly~\cite{adamsOverviewDistributedSpacecraft2023, cramerDesignTestingAutonomous2021, millerStarlingCubeSatSwarm2024, loweConceptOperationsSWARMEX2024, chengEagleEyeNanosatelliteConstellation2024}. A generalized framework is necessary for creating reusable solutions, preventing mission and space application designers from needing to re-implement and engineer the same tools or components. Steps are being taken towards creating flight software solutions tailored specifically towards satellite swarm missions, notably the efforts of the French Space Agency (CNES) to enable swarm missions through its KOSMOS flight software framework~\cite{alvessuanaSwarmSoftware2025, lambertSwarm2026}. For such frameworks and systems to become a reality, the fundamental question of how these systems can be interacted with and operated as a single entity needs to be answered.

The answer to this question lies in the algorithms and methods produced by the field of distributed computing. Distributed computing solves problems where each distributed entity only has a partial knowledge of the many parameters required to solve a problem. The main difficulty lies in the entities cooperating in the achievement of a common goal without the instantaneous knowledge of the current state of all other entities \cite{raynalDistributedAlgorithmsMessagePassing2013}. This mirrors the problem faced by DSS and satellite swarm implementations. 

Within this context, distributed applications first require a way to share information and data, followed by a means of distributed control and decision making before any mission or science objective can be considered. This paper will focus on implementing a strongly eventually consistent distributed state service with the goal of supporting subsequent capabilities which will depend on its information sharing functions. The various functions of Space Situational Awareness (SSA), defined as the detection, tracking, and cataloging of objects in orbit, provide a useful target domain for validating distributed space application solutions and will be used to evaluate the proposed distributed state service.

Therefore the main contributions of this work are as follows:
\begin{itemize}
    \item A characterization of the key capabilities and enabling technologies required for DSS to be implemented.
    \item The introduction of a distributed computing framework to enable the development of distributed applications for satellite swarms.
    \item The design, development and evaluation of a strongly eventually consistent state service for SSA catalog sharing, which reduces the reliance on ground-to-satellite communication.
\end{itemize}

Section~\ref{sec:background} will present the background and a brief motivation behind this work. The proposed framework and distributed state design will be presented in Sections~\ref{sec:design} and \ref{sec:diststate}, respectively. Section~\ref{sec:method} will present the methodology of this paper, while Section~\ref{sec:results} provides the results with discussion. A conclusion is drawn in Section~\ref{sec:conclusion}.

\section{Background and Motivation}
\label{sec:background}
To develop a framework for reliable distributed computing for satellite swarms, DSS first needs to be defined and characterized. This can be done in a number of ways, such as by their capabilities or enabling technologies. This knowledge can then be used to identify the foundations required for designing distributed applications for satellite swarms.

The DARPA OFFensive Swarm-Enabled Tactics (OFFSET) program~\cite{chungOffensive2017} identified five parameters to describe capability gaps in the realization of swarms in general:
\begin{enumerate}
    \item Number: The number of satellites within a swarm.
    \item Collective Complexity: The complexity that emerges from the interactions between swarm members.
    \item Human-swarm Interaction: The ease and way in which operators direct the actions of the swarm.
    \item Heterogeneity: The diversity of platforms and roles available in the swarm.
    \item Agent Complexity: The complexity of each individual member in the swarm.
\end{enumerate}
These parameters are applicable to satellite swarms and DSS. The gaps in capabilities for a given DSS can be formulated based on the parameters listed above, making it easier to identify solutions capable of filling these gaps and ensuring mission success.

With these capability gaps in mind, NASA Ames attempts to characterize DSS based on a number of enabling technologies that build on top of each other~\cite{cramerDesignTestingAutonomous2021}. The enabling technologies are defined as follows:
\begin{itemize}
    \item Network: The ability to pass messages between spacecraft in a DSS.
    \item Distributed State: The shared knowledge of the state of the DSS.
    \item Command and Control: The ability to control the DSS as a single entity.
    \item Planning: Utilizing available information to create DSS-level commands.
    \item Science: An autonomous DSS can dynamically balance and re-assign tasks without the intervention of the ground.
\end{itemize}
First, a communication network is required that will allow a distributed state to be shared. This enables the command and control of the DSS, and informs the way in which plans are made. All of this creates a system on top of which science can be performed~\cite{cramerDesignTestingAutonomous2021}.

Distributed computing can provide solutions which can operate within these different technology layers and fill the swarm capability gaps, an approach not taken in previous work. The nature of DSS and satellite swarms results in a system filled with uncertainty created by various factors such as asynchrony, multiplicity of control flows, absence of shared memory and global time, failure, dynamicity, mobility, and so on. Distributed computing can be characterized by the term uncertainty and aims to provide solutions around it~\cite{raynalDistributedAlgorithmsMessagePassing2013}.

When considering distributed services, the CAP theorem introduces a trade-off between consistency, availability and partition-tolerances~\cite{gilbertBrewersConjectureFeasibility2002}. It is possible to design services which guarantee only two of these qualities. The qualities most important to a DSS will depend on its mission and requirements. However, generally speaking, the space environment and swarm dynamics will cause interruptions to communication between satellites~\cite{akopyanNetworkCharacterizationNanoSatellite2023}, or even the loss of satellites, making partition tolerance essential. This means that the CAP theorem trade-off for DSS will mainly be between availability and consistency. This work will focus on data availability since it is a generally desirable quality and data synchronization is costly.

\section{Framework Design}
\label{sec:design}
\begin{figure}[htbp]
\centerline{\includegraphics[width=0.65\linewidth]{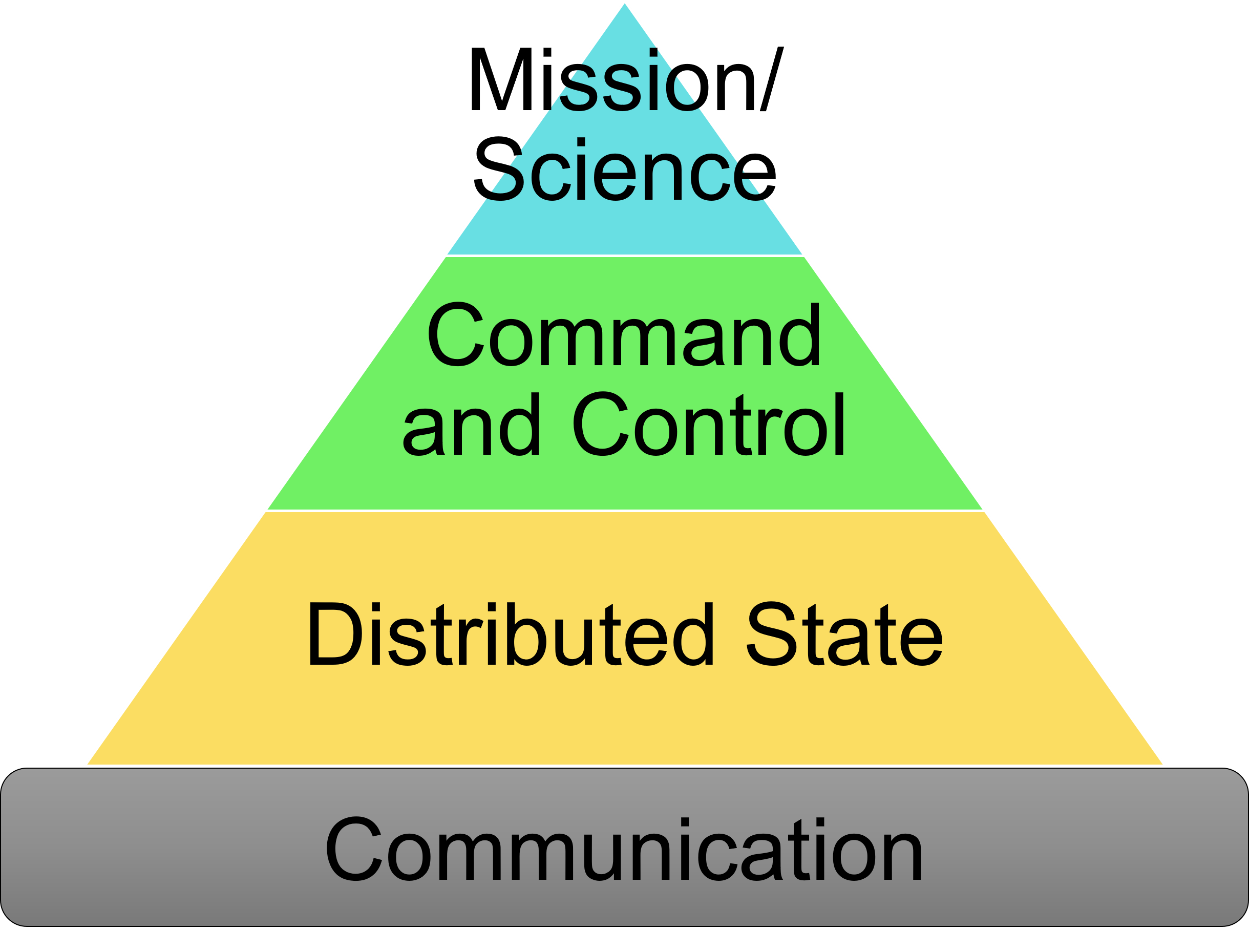}}
\caption{Conceptual framework for distributed space applications.}
\label{fig:distapplayer}
\end{figure}

From an application perspective, the capability gaps and enabling technologies mentioned in the previous section, can be combined and condensed into three layers, forming a conceptual framework for distributed space applications built on top of a foundation of communication. Fig.~\ref{fig:distapplayer} illustrates this conceptual framework. Distributed algorithms and methods must be selected to fulfill these layers.

The major assumption of this framework is that a means of communication and way to pass messages exists between every satellite in a swarm. This is why communication is presented as the foundation on top of which distributed applications are built. Whether messages arrive eventually, or not at all, will depend on the exact characteristics of the DSS. Some communication losses are naturally to be expected in a space environment.

At the base of this framework lies the distributed state. To complete any sort of operation or computation across a swarm of satellites, information and data must be reliably shared and replicated. Without sufficient information about the state of the swarm, reliable control and planning of the system is impossible. This makes a reliable and fault-tolerant distributed state essential for any distributed space application.

Building on top of this is command and control. Services that fall within this layer allow a swarm of satellites to be controlled as a single entity. This means that, given a task by a human operator, the swarm will determine the best way to accomplish this task, considering its current state and configuration. This layer provides the autonomy required to feasibly operate a swarm of satellites, including techniques that facilitate the selection of suitable leaders, creating mission plans from received commands, scheduling, and so on.

Once these two foundations are in place, missions can be designed and science completed. Distributed applications must consider the constraints imposed by the characteristics and design of the DSS itself in conjunction with the needs and requirements of the mission.

\section{Strongly Eventually Consistent Distributed State}
\label{sec:diststate}
The first step in creating fault-tolerant and scalable distributed applications is the implementation of a reliable distributed state mechanism. For this reason, one of the major contributions of this paper is the design and validation of a strongly eventually consistent distributed state.

The choice of strong eventual consistency (SEC) was made with the constraints of satellite swarms in mind. The space environment requires services resilient to the partitions that will inevitably occur in a swarm network~\cite{akopyanNetworkCharacterizationNanoSatellite2023}. For swarm services and applications to remain available, a trade-off between consistency and partition tolerance is imposed by the CAP theorem~\cite{gilbertBrewersConjectureFeasibility2002}. Eventual consistency provides an alternative approach by not requiring nodes to synchronize with other nodes. Instead, operations are performed asynchronously and in possibly different orders~\cite{shapiroComprehensiveStudyConvergent2011}. To achieve eventual consistency the following is required: eventual delivery of all updates to all correct nodes, eventual convergence to an equivalent state by all correct nodes, and the termination of execution of all methods. An object is strongly eventually consistent if it is eventually consistent and displays strong convergence, where all nodes that have delivered the same updates have equivalent state~\cite{shapiroCRDT2011}. SEC provides a solution to the CAP theorem, while being acceptable for applications that do not require linearizable sequential consistency, where a strict order of operations must be enforced.

Conflict-free Replicated Data Types (CRDT) is a family of distributed data types that ensures SEC across all nodes in a system~\cite{shapiroCRDT2011}. CRDT operations are sent asynchronously, possibly arriving in different orders, leaving the nodes to reconcile conflicting updates and eventually apply all updates. CRDT is designed around some mathematical properties that ensure strong eventual consistency while not using consensus by design~\cite{shapiroCRDT2011}. These qualities make CRDT well suited for cases where communication between spacecraft is not guaranteed and data is not required to be immediately consistent across all nodes. By design, CRDT are light to implement, requiring few messages and constraints on order of delivery and reliability.

A CRDT register is a memory cell that stores an object supporting the operations \textit{assign}, to update its value, and \textit{value}, to query it~\cite{shapiroComprehensiveStudyConvergent2011}. Non-concurrent \textit{assigns} overwrite earlier updates, while concurrent updates do not commute~\cite{shapiroComprehensiveStudyConvergent2011}. A last-write-wins register (LWW-Register)~\cite{johnsonRFC6771975} associates timestamps with each update to create a total order of assignments, providing safeguards to the non-commutativity of normal registers by having one update take precedence over another~\cite{shapiroComprehensiveStudyConvergent2011}. Timestamps are assumed to be consistent with causal order, totally ordered, and unique. A Lamport clock, for instance, fulfills these requirements. LWW-registers are well suited as a distributed state mechanism for DSS, as they are widely used in distributed systems in applications such as replicated file systems, where the object a register stores may be a file or a block in a file~\cite{shapiroComprehensiveStudyConvergent2011}.

\section{Methodology}
\label{sec:method}
To validate the use of CRDT to create a reliable distributed state, a LWW-Register-based key-value store was implemented to fulfill the task of distributing a SSA catalog across a swarm of satellites. The catalog is represented as a list of JSON objects, generated by CelesTrak, containing the general perturbation (GP) data of 5 different orbital objects over a one year period.

\subsection{Test Environment}
\begin{figure}[htbp]
\centerline{\includegraphics[width=0.9\linewidth]{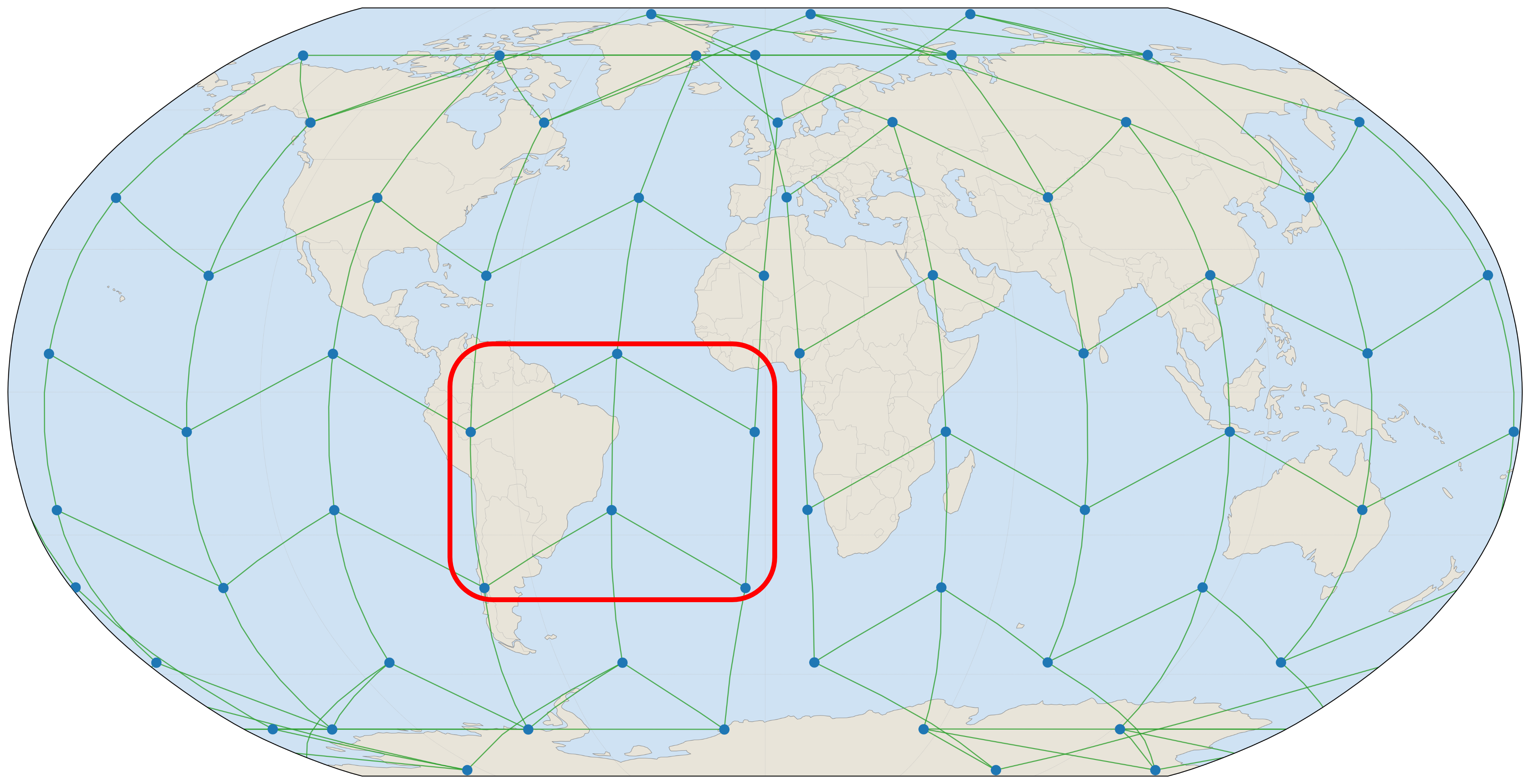}}
\caption{Wireless communication topology of the 66-Satellite constellation.}
\label{fig:sattopology}
\end{figure}

Tests are performed on a simulated 66-satellite constellation based on the Iridium constellation. The constellation, a Walker Star type~\cite{walkerConstellations1984}, is comprised of 6 orbital planes at an 86.4 degree inclination, containing 11 satellites each at 14.36 mean revolutions per day. Each satellite is connected to a maximum of 4 adjacent nodes. The network topology was generated using SatComTopology and SatGoNetEm~\cite{ariassuarez2026satgonetem}. Fig.~\ref{fig:sattopology} presents a visual representation of the 66-satellite constellation, where satellites are represented as blue dots and inter-satellite links (ISL) are included as green lines.

Simulations of the static topology are run using the GoNetEm network emulator~\cite{gonetem}. GoNetEm emulates nodes, in this case satellites, using Docker containers and emulates the network connections between them. The containers provide a Linux environment allowing prototypes to be rapidly built for validation using any programming language of choice. Users have the ability to define network connections and link characteristics.

The wireless network communication was configured to match the topology presented in Fig.~\ref{fig:sattopology}. Additionally, each node was given an extra link to a special ground station (GS) node. Each link is accessible by a virtual Ethernet interface on each node. To emulate ISL conditions, the links were configured to have a maximum bandwidth of 200Kbps, 250ms of latency (125ms per interface) with 25ms of jitter, and 0.001\% of loss. The bandwidth was set following CNES's radio-frequency ISL expectations and the delay accounts for both transmission delay and delay resulting from the satellite system itself. The GS links were configured identically.

It is assumed that GS contacts are costly for the satellites in terms of power and resources, and are not completely reliable. GS messages travel long distances and are prone to atmospheric disturbances and GS tracking errors. Therefore, any reduction in the number of GS message exchanges is seen as a positive. In comparison, ISL messages are seen as more reliable as relative positions between satellites are fixed and less power is required to propagate messages given the distance between satellites and lack of atmosphere.

\subsection{Implementation}
The pseudocode specification for an operation-based LWW-Register CRDT from \cite{shapiroComprehensiveStudyConvergent2011} was implemented in Python 3 and coupled with a dictionary to create a key-value store. An object ID from the catalog data points to a LWW-Register object allowing updates or queries of the latest version of the GP data received for that object. CRDT update messages are sent encapsulated in UDP packets. Each update with a specific object ID contains a distinct snapshot of the GP data for the same object, and every update with the same timestamp from the same origin is identical. The LWW-Register applies an update if the incoming timestamp is greater than or equal to the stored one. Since updates originate only from the ground, total ordering is trivial and causal consistency reduces to "highest timestamp wins." An update message is forwarded if and only if it updates the local state, reducing the strain on the ISL network.

\begin{figure}[htbp]
\centerline{\includegraphics[width=0.85\linewidth]{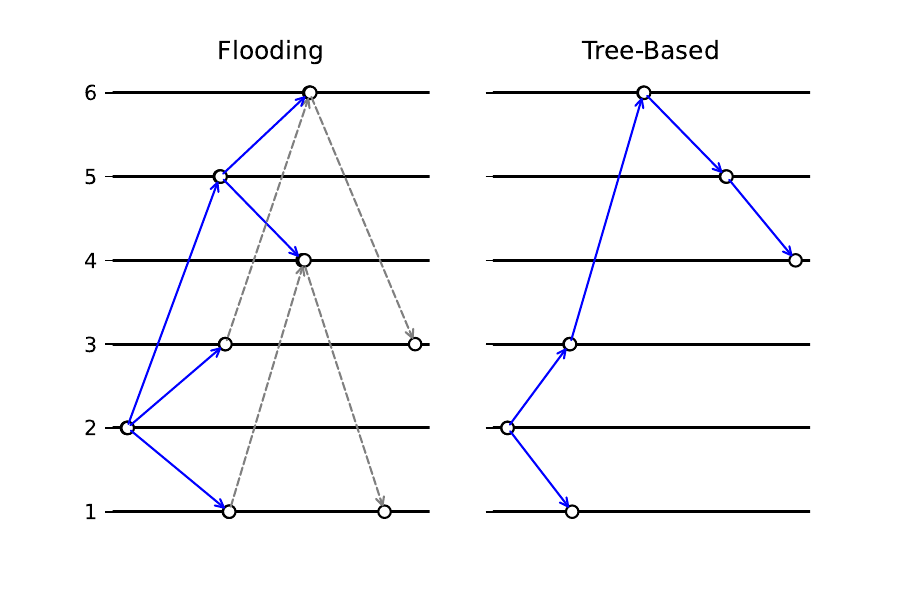}}
\caption{Message passing events for the dissemination of a single update.}
\label{fig:chronogram_comparison}
\end{figure}

Fig.~\ref{fig:chronogram_comparison} presents how messages are passed between satellites for a single update through a 6-satellite subset (indicated by the red rectangle in Fig.~\ref{fig:sattopology}) of the 66-satellite constellation for flooding and tree traversal dissemination. New messages, not already received by a node, are represented by blue arrows, while redundant, already received, messages are represented by gray dashed arrows. Messages represented by the gray dashed arrows are dumped upon reception. 

\subsection{Single-update Test Scenario}
In this scenario a single update to the GP data of a tracked space object is made on the ground and must be propagated to the entire 66-satellite constellation. Validation is performed using three variations of this scenario, the first uses only ground-to-satellite communication, while the other two use the CRDT-based distributed state coupled with different types of dissemination over the ISL network. The number and type of messages will be tracked during the test. 

The three scenario variations are as follows:
\paragraph{Direct Uplink (No ISL)} This case acts as the baseline comparison modeled after the "bent-pipe" architecture seen in current satellite constellations~\cite{denbyOrbitalEdgeComputing2020}. Communication occurs only between the ground station and each individual satellite. No ISL communication occurs between satellites. An arbitrary time, $t$, passes between ground-to-satellite contacts. In the real world, this would be the amount of time for the ground station to contact the next satellite based on how they are orbiting the Earth. For this test, $t$ was set to 5 seconds.

\paragraph{CRDT + Flooding} An update is first sent from the ground to a single satellite. An update is performed and disseminated via the ISL network to the entire swarm via a flooding mechanism. Messages are forwarded to every neighbor via the links available to the satellite, excluding the link from which the message was originally received. If a message has not yet been received, the satellite forwards it, otherwise the message is dropped. Flooding provides a simple form of reliable broadcast. A new update will eventually arrive to every satellite provided no network partitions exist.

\paragraph{CRDT + Tree Traversal} Once again, an update is first sent from the ground to a single satellite. An update is performed and then disseminated via the ISL network to the entire swarm via a mechanism which applies tree traversal. Messages traverse a spanning tree to reach every satellite in the swarm. The spanning tree is constructed using a simple algorithm that, when initiated, causes a satellite to send a query to each of its neighbors. If a query is received and the receiving satellite has no parent, the sending satellite is assigned as its parent and the receiving satellite then sends a query to all of its neighbors. In the current implementation, only satellite 1 initiates this process at the start of the test.

\subsection{Multi-update Test Scenario}
In this scenario multiple updates to the same tracked object are issued to one, five, ten, and twenty satellites spread across the network. Updates are made sequentially, with no delay between sends and incrementing version numbers, resulting in multiple updates propagating concurrently from different entry points into the ISL network. Tree traversal is used for message passing. This scenario aims to highlight the benefit of this CRDT-based approach when multiple updates are introduced.

\section{Results and Discussion}
\label{sec:results}
This section presents the results of the validation tests performed and discusses the outcome. Table~\ref{tab:nmessage} presents the number of messages passed for each variant of the single-update scenario. Messages are classified as either GS or ISL.

\begin{table}[htbp]
\caption{Number of Messages for a Single Catalog Update}
\begin{center}
\begin{tabular}{|c|c|c|c|}
\hline
\textbf{Test}&\textbf{GS}&\textbf{ISL}&\textbf{Total} \\
\textbf{Case}&\textbf{Messages}&\textbf{Messages}& \\
\hline
Direct Uplink (No ISL)&66&0&66 \\
CRDT + Flooding&1&177&178 \\
CRDT + Tree Traversal&1&65&66 \\
\hline
\end{tabular}
\label{tab:nmessage}
\end{center}
\end{table}

The table shows that a CRDT approach is effective in reducing the number of ground station messages required for a single catalog update. Instead of updating every satellite in the constellation from the ground, only one needs to be updated for the new data to propagate across the entire swarm.

The results show that the CRDT-based distributed state is effective regardless of the dissemination mechanism used and that this mechanism is what determines the number of ISL messages needed to complete an update. Network flooding is known to send $2E-N+1$ messages, where $E$ is the number of edges, in this case ISL links, and $N$ is the number of nodes, or satellites. Meanwhile, dissemination using tree traversal sends $N-1$ messages to fully propagate a message. The fact that the experimental results match the expected analytical outcome, shows that these mechanisms have been applied correctly and can be operated in a satellite-like environment.

Initially, 354 messages were exchanged to construct the spanning tree used by the tree traversal mechanism. This number will increase if a more complex or robust algorithm were to be implemented. The number of messages sent during tree construction are ultimately not included in the results table as the spanning tree is only constructed once at startup, provided the network topology does not change. Should a change occur, the spanning tree will be rebuilt, incurring an additional cost of 354 messages to complete the process. While ISL connectivity is time-varying for low-earth-orbit constellations, the expected variability under nominal operations is low.

\begin{figure}[htbp]
\centerline{\includegraphics[width=0.9\linewidth]{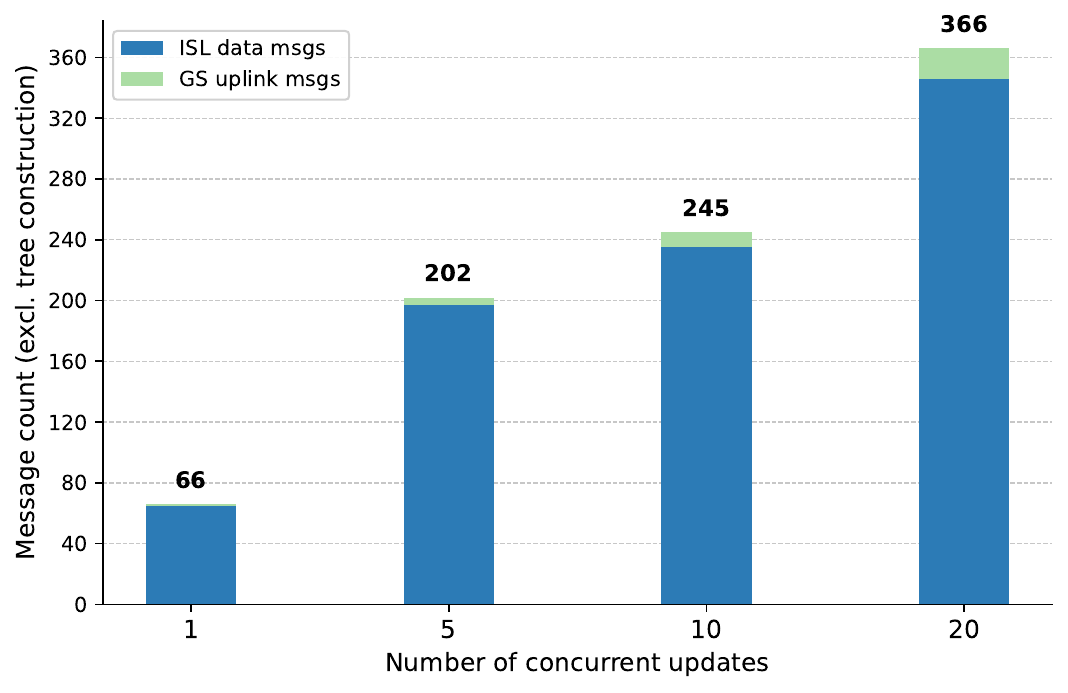}}
\caption{Messages generated under concurrent updates.}
\label{fig:concupdate}
\end{figure}

Fig.~\ref{fig:concupdate} presents the number of messages sent to complete the specified number of catalog updates for the same tracked object for the multi-update scenario (excluding the messages sent during initial tree construction). The graph exhibits sub-linear growth as the number of updates increases, since nodes already holding a newer version will drop outdated update messages upon reception. This behavior emerges when updates are injected faster than the network can fully propagate them, since a node that has already received a newer update will discard any older one. Should enough time pass between updates, a single new update will propagate through the entire network before the next one is uplinked from the ground. Therefore, the single-update result serves as an upper bound on the number of messages required per update.

In the case of LWW-Registers, consistency will be reached if the latest update arrives at all nodes. In this way, the distributed state is "self-healing" and can recover from failures and partitions within the swarm, as the latest update overrides any earlier update for the same tracked object. If the satellite network itself provides delivery or timeliness guarantees for message delivery, then these will be inherited by a CRDT-based distributed state.

One practical benefit of this distributed state approach is the real-world time savings. Directly up-linking updates to all satellites requires more time due to the limited windows of ground contact caused by satellite orbital periods. By requiring only a single up-link per update, and since satellites can continuously communicate with each other via ISL, a distributed approach completes dissemination significantly faster, even accounting for propagation delay within the satellite network. In the context of this test, with ground contact intervals set to 5 seconds, direct up-link to all satellites completed after 330 seconds, while the distributed cases completed within 5 seconds. With real orbital mechanics, ground contact intervals are considerably longer, making the advantage of the distributed approach even more pronounced.

A further practical benefit arises from the cost of ground-to-satellite communication in terms of power consumption and reliability. Reducing the number of required ground contacts directly lowers overall power consumption and improves the robustness of mission data dissemination.

With regards to future work, distributed consensus should be implemented and tested for cases requiring linearizable sequential consistency, such as applications that modify real-world properties of the satellite swarm or require operations to arrive in order for safe execution. Services in subsequent framework layers should then be investigated and implemented, enabling satellite swarms to take on full missions and complete real science objectives. An important avenue for validating this approach is testing on real satellite software and eventually hardware, in a dynamic orbital environment. CNES is currently building the simulation environment, software, and hardware platforms on which future work will be tested~\cite{alvessuanaSwarmSoftware2025, lambertSwarm2026}. Finally, comparison to existing swarm approaches, such as those found in NASA's Starling mission~\cite{millerStarlingCubeSatSwarm2024}, should be performed to evaluate the full benefit of a distributed computing approach to satellite swarm and DSS flight software.

\section{Conclusion}
\label{sec:conclusion}
The rise of DSS and large constellations has necessitated a framework to enable fault-tolerant and reliable distributed space applications. This paper presented a conceptual distributed space application framework, covering distributed state, command and control, and scientific mission, demonstrating its potential through the implementation and validation of the distributed state layer.

A strongly eventually consistent distributed state based on CRDT was implemented and evaluated in the context of SSA catalog dissemination. The distributed approach required fewer ground contacts and less time to disseminate updates compared to direct ground uplinks. When multiple updates were issued in quick succession, a reduced number of ISL messages were required to converge to the latest version across all satellites.

These results demonstrate that a CRDT-based distributed state is a viable approach to mission data dissemination in satellite swarms, offering meaningful reductions in both ground contact requirements and ISL traffic. The framework presented provides a foundation for building fault-tolerant distributed space applications, with the distributed state layer serving as a building block for the higher-level services that follow.


\bibliographystyle{IEEEtran}
\bibliography{project}

\end{document}